\documentclass[aps, prd, amsmath, floats, floatfix, twocolumn, nofootinbib, showpacs]{revtex4-1}
\usepackage{graphicx}
\usepackage{color}

\newcommand{\be}{\begin{eqnarray}}
\newcommand{\ee}{\end{eqnarray}}

\begin{document}

\title{Note on the Cardoso-Pani-Rico parametrization to test the Kerr black hole hypothesis}

\author{Cosimo Bambi}
\email{bambi@fudan.edu.cn}

\affiliation{Center for Field Theory and Particle Physics and Department of Physics, 
Fudan University, 200433 Shanghai, China}

\date{\today}

\begin{abstract}
The construction of a generic parametrization to describe the spacetime geometry 
around astrophysical black hole candidates is an important step to test the Kerr 
black hole hypothesis. In the last few years, the Johannsen-Psaltis metric has been
the most common framework to study possible deviations from the Kerr solution 
with present and near future observations. Recently, Cardoso, Pani and Rico 
have proposed a more general parametrization. The aim of the present paper is 
to study this new metric in a specific context, namely the thermal spectrum of 
geometrically thin and optically thick accretion disks. The most relevant finding is 
that the spacetime geometry around objects that look like very fast-rotating Kerr 
black holes may still have large deviations from the Kerr solution. This is not 
the case with the Johannsen-Psaltis metric, which means the latter is missing
an important class of non-Kerr spacetimes.
\end{abstract}


\maketitle


Today, general relativity is quite well tested in the Earth gravitational field, in 
the Solar System, and by studying the orbital motion of pulsars in binary 
systems~\cite{will}. The validity of the theory in more extreme situations has 
still to be verified. For instance, the observed accelerated expansion of the 
Universe may indicate a breakdown of general relativity at large scales~\cite{cosmo}. 
The strong field regime is another unexplored area and today there is an 
increasing interest in the possibility of testing the actual nature of astrophysical 
black hole (BH) candidates~\cite{bh}. In classical general relativity, astrophysical 
BHs should be well described by the Kerr solution; for instance, the presence 
of accretion disks is usually completely negligible~\cite{disk}. However, for 
the time being astrophysical BH candidates are just objects too heavy and
compact to be neutron stars or clusters of neutron stars~\cite{mass}. There is
no evidence that the spacetime geometry around them is described by the 
Kerr metric. We rely on classical general relativity and we believe in the Kerr 
BH hypothesis. However, there are some novel theoretical arguments 
suggesting the possibility of macroscopic deviations from classical predictions 
around BHs~\cite{gg}.

The Kerr nature of astrophysical BH candidates can be potentially tested by 
studying the properties of the radiation emitted by the gas in the accretion disk. 
With this approach, one can check if the metric around the compact object is 
described by the Kerr solution, which determines the motion of the gas in the 
accretion disk and the propagation of the radiation from the disk to the distant 
observer, while it is not possible to distinguish Kerr BHs of general relativity 
from Kerr BHs in alternative theories of gravity~\cite{kerr}. The strategy is then 
to use a framework similar to the PPN (Parametrized Post-Newtonian) 
formalism~\cite{will}, which has been very successful for tests of general relativity
in the Solar System. The idea is to have a very general metric with a number 
of ``deformation parameters'' that measure possible deviations from the 
Kerr geometry. Like the traditional $\beta$ and $\gamma$ of the PPN formalism, 
the value of the deformation parameters has to be determined from observations, 
and {\it a posteriori} one can verify if the measurements are consistent with the 
predictions of general relativity. If we had a very general formalism, such a 
test-metric should be able to reduce to any metric describing the gravitational
field around a compact object in any alternative theory of gravity for a suitable 
choice of the value of the deformation parameters. Unfortunately, such an 
approach in the strong field limit is much more complicated than its counterpart 
in the weak field regime, and today there is not yet a completely satisfactory 
formalism to test the Kerr BH hypothesis.

{\it New parametrization ---}
In the last few years, the most popular parametrization to test the Kerr nature of 
astrophysical BH candidates has been the Johannsen-Psaltis (JP) metric~\cite{jp}. 
Such a metric has an infinite number of deformation parameters and reduces to 
the Kerr solution when all the deformation parameters are set to zero. Recently, 
Cardoso, Pani and Rico suggested a generalization of the JP parameterization. 
In Boyer-Lindquist coordinates, the Cardoso-Pani-Rico (CPR) parametrization 
reads~\cite{cpr}
\begin{widetext}
\be
ds^2 &=& - \left(1 - \frac{2 M r}{\Sigma}\right)\left(1 + h^t\right) dt^2
- 2 a \sin^2\theta \left[\sqrt{\left(1 + h^t\right)\left(1 + h^r\right)} 
- \left(1 - \frac{2 M r}{\Sigma}\right)\left(1 + h^t\right)\right] dt d\phi \nonumber\\ 
&& + \frac{\Sigma \left(1 + h^r\right)}{\Delta + h^r a^2 \sin^2\theta} dr^2
+ \Sigma d\theta^2 
+ \sin^2\theta \left\{\Sigma + a^2 \sin^2\theta \left[ 2 \sqrt{\left(1 + h^t\right)
\left(1 + h^r\right)} - \left(1 - \frac{2 M r}{\Sigma}\right)
\left(1 + h^t\right)\right]\right\} d\phi^2 \, , \quad
\ee
where $M$ is the BH mass, $a = J/M$ the BH spin parameter, $J$ the BH spin 
angular momentum, $\Sigma = r^2 + a^2 \cos^2 \theta$, $\Delta = r^2 - 2 M r + a^2$, and
\be
h^t = \sum_{k=0}^{+\infty} \left(\epsilon_{2k}^t 
+ \epsilon_{2k+1}^t \frac{M r}{\Sigma}\right)\left(\frac{M^2}{\Sigma}
\right)^k\, , 
\qquad
h^r = \sum_{k=0}^{+\infty} \left(\epsilon_{2k}^r
+ \epsilon_{2k+1}^r \frac{M r}{\Sigma}\right)\left(\frac{M^2}{\Sigma}
\right)^k \, .
\ee
\end{widetext}
Here we have two infinite sets of deformation parameters, namely $\{\epsilon_k^t\}$ 
and $\{\epsilon_k^r\}$.
The CPR parametrization reduces to the JP one for $h^t = h^r$ and to the 
Kerr solution when $h^t = h^r = 0$. The aim of the present paper is to figure 
out the possible advantages, if any, of the new metric to test the Kerr BH hypothesis.

{\it Constraints ---}
For the time being, the most trustworthy technique to probe the spacetime 
geometry around astrophysical BH candidates is likely the so-called continuum-fitting 
method; that is, the study of the thermal spectrum of a geometrically thin and 
optically thick accretion disk~\cite{cfm}. While it has been developed to 
measure the spin parameter of BH candidates under the assumption of the 
Kerr background, this technique can be easily extended to test the Kerr BH 
hypothesis~\cite{cfm2}. With already available X-ray data, the nature of 
astrophysical BH candidates can be investigated even with the analysis of the 
iron K$\alpha$ line~\cite{iron}, but the fact that the model has several parameters  
to be measured during the fitting procedure makes this technique more tricky,
and at present it can just be used to rule out some very exotic spacetimes with
qualitatively different iron lines~\cite{iron2}. Other approaches, like the study
of QPOs~\cite{qpo} or the estimate of the jet power~\cite{jets}, are model-dependent
and not yet mature to test fundamental physics, while high resolution observations
of the accretion flow around the supermassive BH candidate in the Milky Way are 
not yet available~\cite{shadow}.

At present, there are 10~stellar-mass BH candidates for which the current data allow us
to get reliable constraints from the continuum-fitting method. In case we assume 
that the spacetime around these objects is described by the Kerr solution, it is 
possible to estimate the spin parameter. If we relax the Kerr BH hypothesis, it is 
usually possible to constrain some combination of the spin and of possible deviations 
from the Kerr solution, because of the strong correlation between them. Recently, 
the measurements of these 10~stellar-mass BH candidates have been reconsidered 
in the JP framework and the constraints on the dimensionless spin parameter 
$a_* = a/M$ and JP deformation parameter $\epsilon_3$ have been 
obtained~\cite{cfm3}\footnote{In the JP metric, $\epsilon_0 = 0$ is required by 
asymptotic flatness, while $\epsilon_1$ and $\epsilon_2$ must be small to satisfy 
the Solar System tests~\cite{cpr}. $\epsilon_3$ is the first deformation parameter 
without constraints and there are no qualitative differences between $\epsilon_3$ 
and higher order terms~\cite{agn}.}. Here we want to use the same procedure to 
the CPR parametrization to figure out possible differences and advantages.

\begin{figure*}
\begin{center}
\includegraphics[height=6cm]{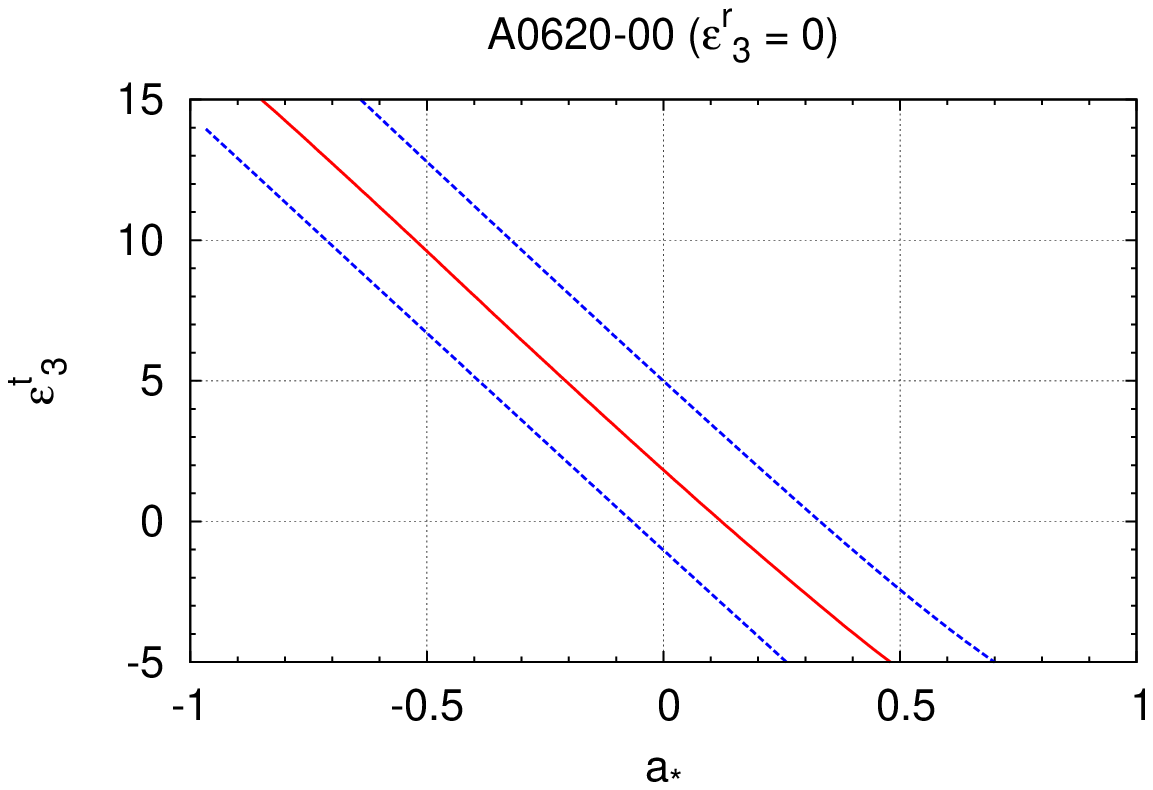}
\includegraphics[height=6cm]{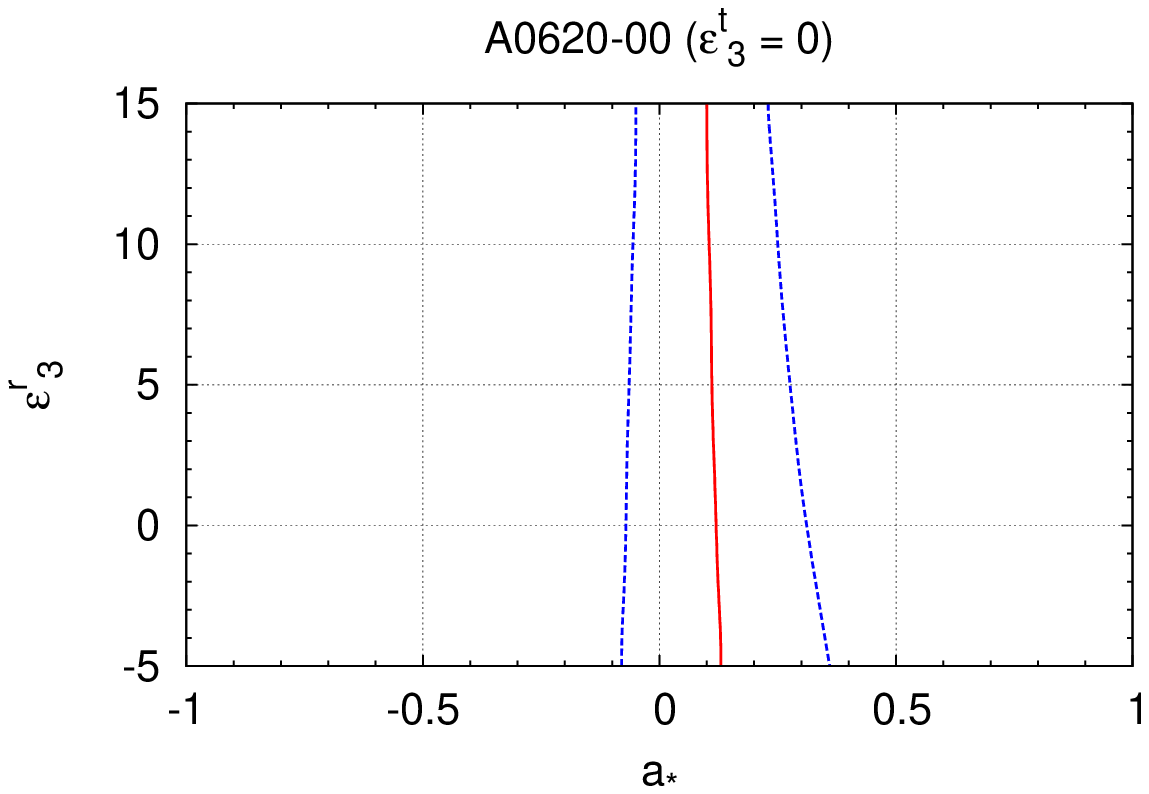}
\end{center}
\vspace{-0.8cm}
\caption{Disk thermal spectrum constraints on possible deviations from the Kerr 
geometry in the spacetime around the BH candidate in A0620-00. Left panel: 
constraints on $a_*$ and $\epsilon_3^t$ assuming that all the other deformation 
parameters vanish. Right panel: constraints on $a_*$ and $\epsilon_3^r$ assuming 
that all the other deformation parameters vanish. The red solid line indicates the
spacetimes that, for a fixed deformation parameter, have the disk's spectrum more 
similar to the one of a Kerr BH with spin $a_* = 0.12$. The blue dashed lines 
are the $1\sigma$ boundary. \label{f-a0620}}
\vspace{0.3cm}
\begin{center}
\includegraphics[height=6cm]{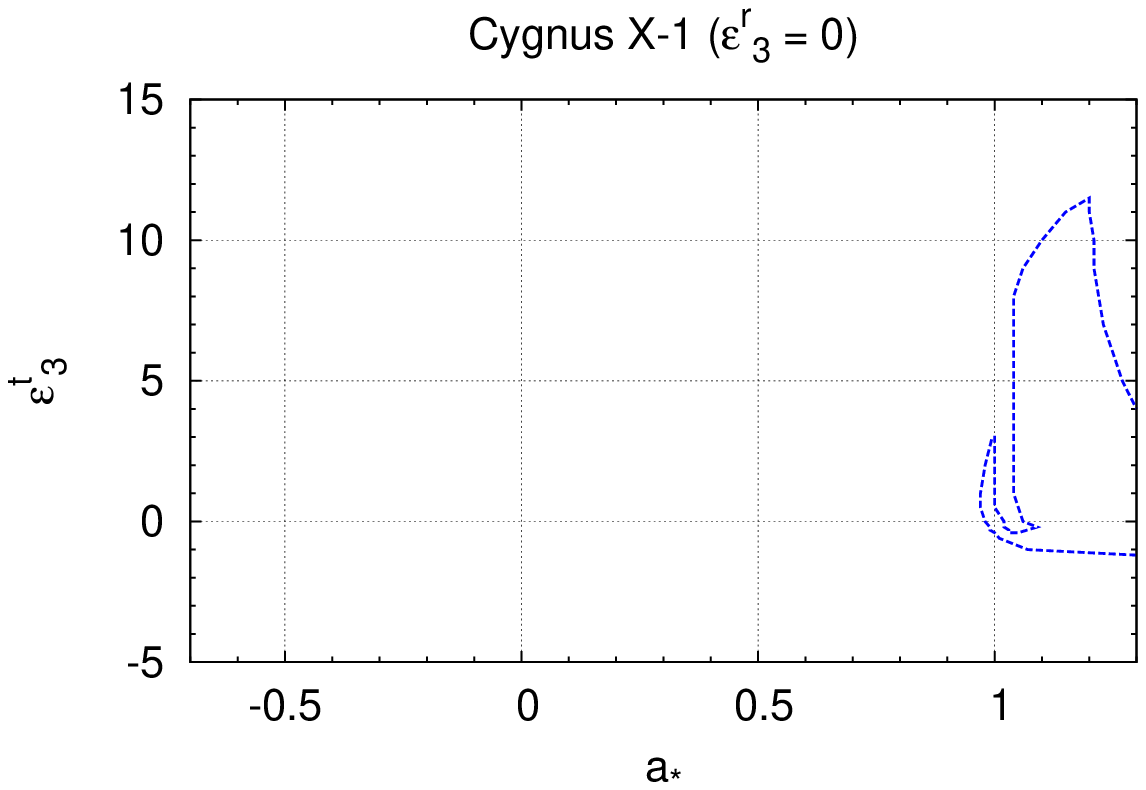}
\includegraphics[height=6cm]{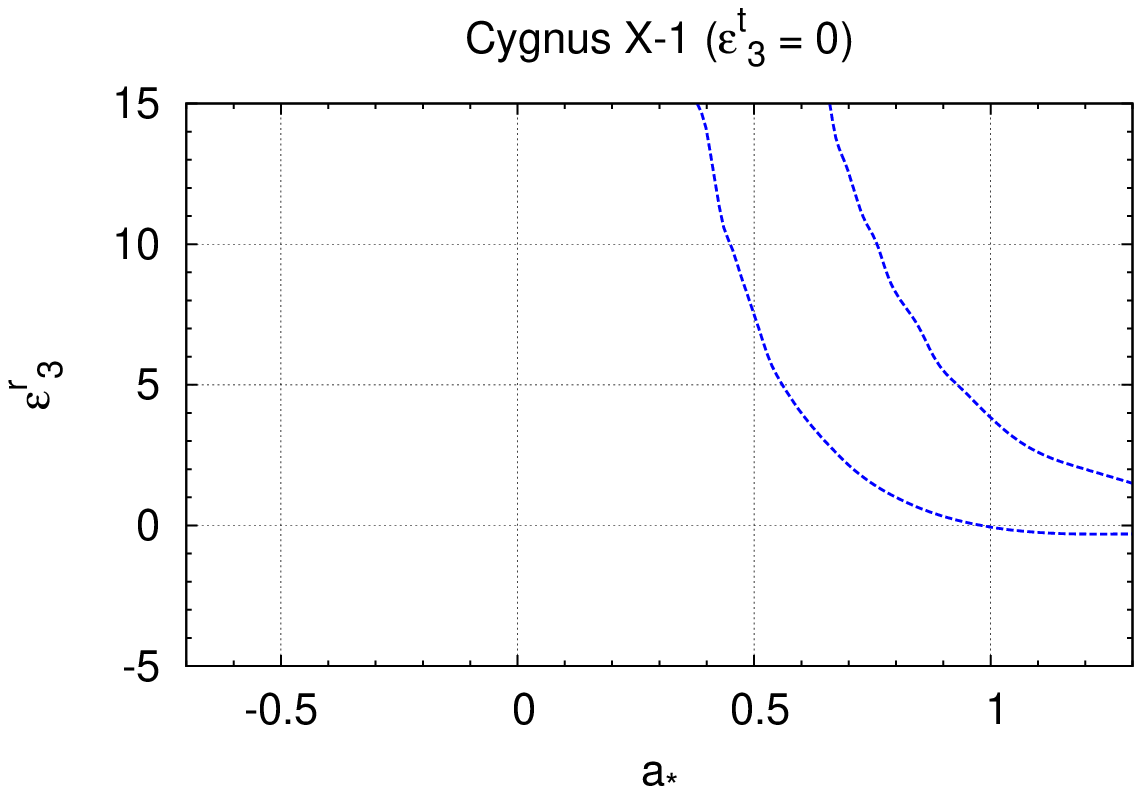}
\end{center}
\vspace{-0.5cm}
\caption{Disk thermal spectrum constraints on possible deviations from the Kerr 
geometry in the spacetime around the BH candidate in Cygnus~X-1. Left panel: 
constraints on $a_*$ and $\epsilon_3^t$ assuming that all the other deformation 
parameters vanish. Right panel: constraints on $a_*$ and $\epsilon_3^r$ assuming 
that all the other deformation parameters vanish. The blue dashed lines are the 
boundaries separating objects that look more like Kerr BHs with $a_* > 0.98$ 
(internal area) and the ones that look more like Kerr BHs with $a_* < 0.98$ 
(external area). \label{f-cygx1}}
\end{figure*}

Instead of reconsidering all of the 10~stellar-mass BH candidates, for the purpose 
of the present paper it is enough to study two extreme cases, namely an object 
that looks like a slow-rotating Kerr BH and one that seems to be a very fast-rotating 
Kerr BH. For the former case, a suitable candidate is A0620-00, whose spin has 
been estimated to be $a_* = 0.12 \pm 0.19$ (at $1\sigma$) under the assumption 
of the Kerr background~\cite{a0620}. Proceeding as described in~\cite{cfm3}, one 
can translate the best value into a line on the spin--deformation parameter 
plane, and the uncertainty into an allowed region. The results of this analysis are 
reported in Fig.~\ref{f-a0620}, where the left panel assumes the CPR 
deformation parameter $\epsilon_3^t$ and all the others are set to zero, while in 
the right panel the only non-vanishing deformation parameter is 
$\epsilon_3^r$\footnote{Actually, the leading order term of $h^r$, which is not 
bounded by Solar System tests, is proportional to $\epsilon_2^r$. Here we 
consider $\epsilon_3^r$ in order to have the same corrections in $h^t$ and $h^r$, 
as well as a simple link to the JP metric, whose unbounded leading order 
correction is the term with $\epsilon_3$ and here it is recovered for 
$\epsilon_3^t = \epsilon_3^r$. Moreover, there are no qualitatively different 
constraints on $\epsilon_2^r$, $\epsilon_3^r$, or higher order deformation 
parameters from thin disks. In the case of the JP parametrization, Figs.~2 and 
4 of Ref.~\cite{agn} show that the effect of any deformation parameter is qualitatively 
the same. That is true even with the CPR parametrization, namely the same 
qualitative conclusions hold for any $\epsilon_n^r$, or $\epsilon_n^t$, if we 
study the constraint for a specific $n$.}.  
As we can see, $a_*$ and $\epsilon_3^t$ are strongly correlated, while there is 
almost no correlation between $a_*$ and $\epsilon_3^r$. Like in the JP parametrization, 
it is not possible to constrain the deformation parameters from an object that 
looks like a slow rotating Kerr BH, because observations cannot rule out large 
deviations from the Kerr solution.

The case of an object that looks like a very fast-rotating Kerr BH is more interesting. 
Within the JP framework, such an object can be used to constrain the deformation 
parameter; that is, it is possible to put an upper and a lower bound on the deformation 
parameter independently of the BH spin~\cite{cfm3}. This is not a peculiar feature of 
the JP background, but it is met even in specific BH solutions~\cite{bardeen}. In this 
case, we consider the BH candidate in Cygnus~X-1, which has been studied in 
Ref.~\cite{cygx1} and found that $a_* > 0.98$ (at $3\sigma$ and assuming the Kerr 
metric). Within the JP parametrization with $\epsilon_3$ as the only non-vanishing 
deformation parameter, one finds the bound $0 \lesssim \epsilon_3 \lesssim 4$~\cite{cfm3}. 
That is interesting because the bound is independent of the BH spin, and it is thus 
possible to say that deviations from the Kerr geometry, if any, cannot be too large. If it 
could be possible to get a stronger bound, say $a_* > 0.99$, the constraint 
on $\epsilon_3$ would become more stringent as well. Fig.~\ref{f-cygx1} shows the 
constraints in the case of the CPR parametrization. In the left panel, it is assumed 
that the only non-vanishing deformation parameter is $\epsilon_3^t$. In the right 
panel, the only non-vanishing deformation parameter is $\epsilon_3^r$.

{\it Comments ---}
For Cygnus~X-1, it seems that $\epsilon_3^t$ can be weakly constrained, while $\epsilon_3^r$ 
cannot be constrained at all, in the sense that observations cannot exclude large 
deviations from the Kerr solution. It is also understandable the constraint area of the 
JP parametrization found in~\cite{cfm3}, which is roughly the overlapping region 
from the constraint on $\epsilon_3^t$ and on $\epsilon_3^r$ (the JP metric with 
$\epsilon_3$ as the only non-vanishing deformation parameter corresponds to the 
case in which $\epsilon_3^t = \epsilon_3^r$ and all the other CPR deformation 
parameters vanish). However, Fig.~\ref{f-cygx1} has been simply obtained by the 
comparison of the disk's thermal spectrum in the corresponding backgrounds. 
Some of them may not be physical. For instance, for some values of the spin and the 
deformation parameters these spacetimes may have no event horizon, may have naked 
singularities, pathological regions with closed time-like curves, etc. However, these 
are test-metrics and we may argue that they hold up to some radius $r_\star$, 
like in the case of exterior solutions of compact objects. In the end, we can only 
probe the spacetime at radii larger than the inner edge of the accretion disk, which 
is supposed to be at the innermost stable circular orbit (ISCO) radius. So, if we 
want to remain as general as possible, we have just to check that the accretion 
disk does not enter any pathological region, but it still makes sense to use this 
background.

\begin{figure*}
\begin{center}
\includegraphics[height=6cm]{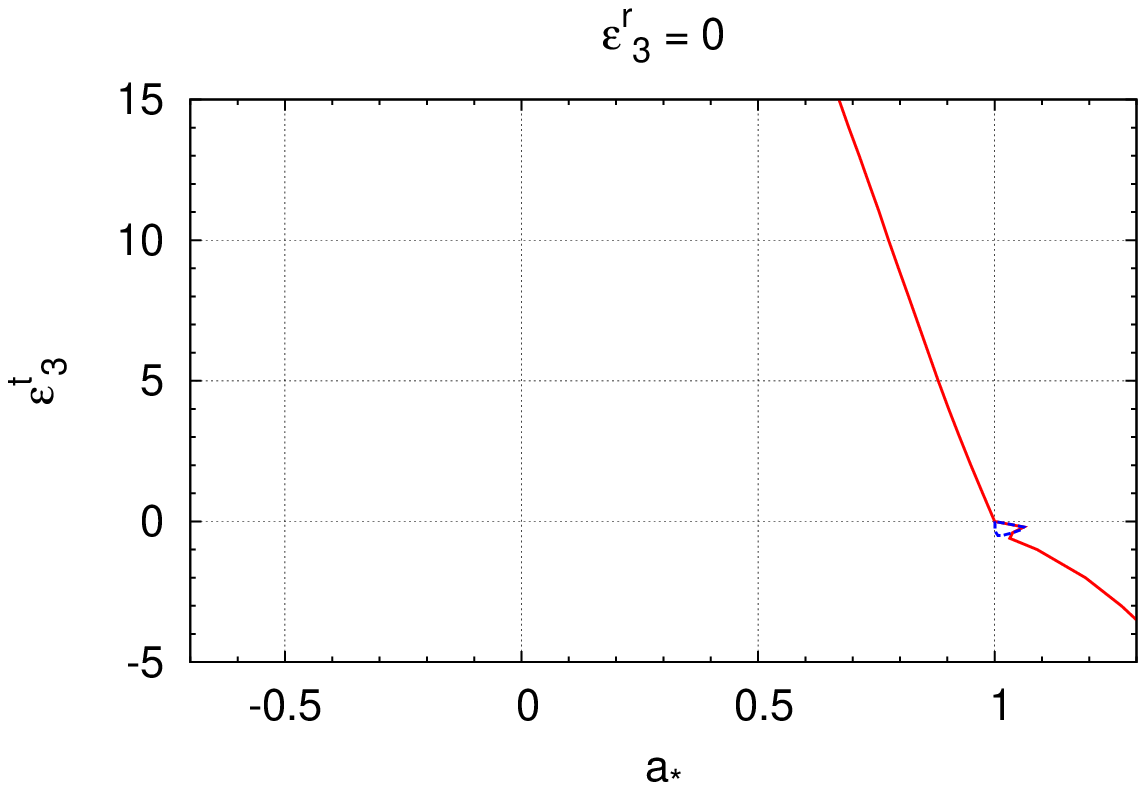}
\includegraphics[height=6cm]{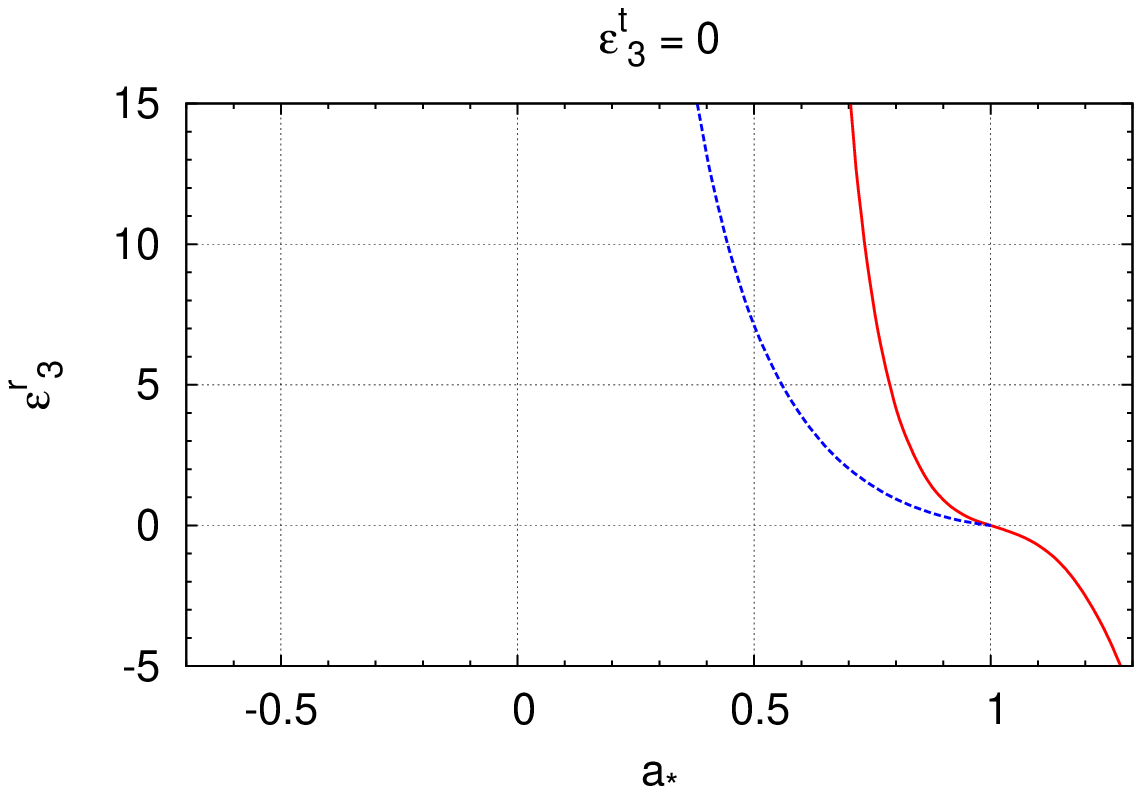}
\end{center}
\vspace{-0.8cm}
\caption{Spin-up/spin-down regions and ISCO stability. Objects on the left side 
of the red solid lines are spun up by a thin accretion disk on the equatorial plane, 
while objects on the right side are spun down. The blue dashed lines separate 
spacetimes in which the ISCO is marginally stable along the radial and the vertical 
direction. Left panel: plane $a_*$ and $\epsilon_3^t$ assuming that all the other 
deformation parameters vanish; the ISCO radius is marginally radially stable 
except in the small area inside the blue dashed closed curve. Right panel: plane 
$a_*$ and $\epsilon_3^r$ assuming that all the other deformation parameters 
vanish; for $\epsilon_3^r > 0$, the ISCO radius is marginally radially stable on 
the left side of the blue dashed line, it is marginally vertically stable on the right 
side; for $\epsilon_3^r < 0$, the ISCO radius is marginally radially stable. \label{f-spin}}
\end{figure*}

However, an important issue concerns the possibility of creating similar
objects. For instance, in the Kerr metric it seems to be impossible to overspin a 
BH up to $a_* > 1$~\cite{ss2}. The simplest and more natural mechanism to spin 
a body up is through accretion from a thin disk. In such a case the evolution of the spin 
is governed by the equation~\cite{spin}
\be\label{eq-spin}
\frac{da_*}{d\ln M} = \frac{1}{M} \frac{L_{\rm ISCO}}{E_{\rm ISCO}} - 2a_* \, ,
\ee
where $E_{\rm ISCO}$ and $L_{\rm ISCO}$ are, respectively, the specific energy 
and the specific angular momentum at the ISCO. 
The central object is spun up/down if the right hand side of Eq.~(\ref{eq-spin}) is
positive/negative and the equilibrium value of the spin parameter is when the
right hand side vanishes. Since $E_{\rm ISCO}$ and $L_{\rm ISCO}$ depend on
$a_*$ and the deformation parameters, for any configuration of the deformation 
parameters there is a different equilibrium spin parameter~\cite{spin}. The equilibrium 
spin parameter for $\epsilon_3^t$ and $\epsilon_3^r$ is shown by a red solid line 
in Fig.~\ref{f-spin}. Objects on the left (right) side of the red solid line are spun up (down)
and it is plausible that objects on the right side can unlikely be created, just as in the 
Kerr case it is apparently impossible to generate a spacetime with $a_* > 1$.
A large fraction of the allowed region in the left panel in Fig.~\ref{f-spin} 
is probably represented by objects with too high spin to be relevant in astrophysics. 
If this is the case, actually the constraint from Cygnus~X-1 would only allow a very 
small range of $\epsilon_3^t$. On the contrary, spacetimes with large $\epsilon_3^r$ 
seem to be possible.

{\it Summary and conclusions ---}
At present, the best strategy to test the Kerr nature of astrophysical BH candidates 
seems to use an approach similar to the PPN formalism, in which a general background 
has a number of free parameters to be measured by observations and one can then 
check if the results are consistent with the predictions of general relativity. For the time 
being, there is not yet a very general parametrization to test the Kerr BH hypothesis. 
In the last few years, most studies have used the JP background~\cite{jp}. Recently, 
Cardoso, Pani and Rico have proposed a more general parametrization~\cite{cpr}. The 
aim of the present paper was to figure out some basic features of the new metric and 
see if it is indeed more useful than the JP parametrization.

The main feature of the CPR parametrization is the presence of two kinds of deformations, 
$h^t$ and $h^r$, which alter respectively the metric coefficients $g_{tt}$ and $g_{rr}$ 
in the non-rotating limit. It turns out that $h^t$ and $h^r$ have quite a different impact 
on possible observables. Here we have just considered the thermal spectrum of thin 
disks, which is today the most reliable technique to probe the spacetime geometry 
around BH candidates. For an object that looks like a non-rotating or a slow-rotating 
Kerr BH, there is a strong correlation between the spin and $h^t$, while there is almost 
no correlation between the spin and $h^r$. Moreover, it seems that $h^r$ cannot be 
constrained at all. In the case of an object that looks like a very-fast rotating Kerr BH, 
$h^t$ can be strongly constrained, independently of the value of the spin parameter.
This was the case of the $h$ in the JP parametrization as well. On the contrary, 
$h^r$ remains difficult to constrain. In other words, even if we observe a BH candidate
that seems to be a very fast-rotating Kerr BH, deviations from the Kerr geometry through 
$h^r$ may be very large. Since this was not the case in the JP parametrization, the latter
was missing an important class of non-Kerr objects and the new parametrization 
may be more convenient for future studies.


{\it Acknowledgments ---}
The author thanks Jiachen Jiang for reading a preliminary 
version of the manuscript and providing useful feedback.
This work was supported by the NSFC grant No.~11305038, 
the Shanghai Municipal Education Commission grant for Innovative 
Programs No.~14ZZ001, the Thousand Young Talents Program, 
and Fudan University.


\end{document}